# The Effect of Plasma Beta on High-*n* Ballooning Stability at Low Magnetic Shear

J W Connor[1,2], C J Ham[1] and R J Hastie[1]

[1] CCFE, Culham Science Centre, Abingdon, Oxon, UK, OX14 3DB

[2] Imperial College of Science and Technology and Medicine, London SW7 2BZ

*Abstract*

An explanation of the observed improvement in H-mode pedestal characteristics with increasing core plasma pressure or poloidal beta, $\beta_{pol}$, as observed in MAST and JET, is sought in terms of the impact of the Shafranov shift, $\Delta'$, on ideal ballooning MHD stability. To illustrate this succinctly, a self-consistent treatment of the low magnetic shear region of the $'s-\alpha'$ stability diagram is presented using the large aspect ratio Shafranov equilibrium, but enhancing both $\alpha$ and $\Delta'$ so that they compete with each other. The method of averaging, valid at low s, is used to simplify the calculation and demonstrates how $\alpha$, $\Delta'$, plasma shaping and 'average favourable curvature' all contribute to stability.

## 1. Introduction

Tokamak performance in H-mode is strongly dependent on the characteristics of the edge pedestal namely its steepness and width, since these determine the effective edge temperature which provides the boundary condition for core transport models [1]. The EPED model [2], which is based on the stability of the edge plasma to ideal MHD peeling-ballooning modes [3, 4], can be used to determine these quantities. There is experimental evidence, e.g. from {DIII-D [5], JT-60U [6], ASDEX Upgrade [7], JET [8] and MAST [9] and, that the pedestal characteristics improve as the core plasma pressure, or poloidal beta, $\beta_{pol}$, increases. This appears to be related to improvements in the ideal MHD stability [10, 11]. The purpose of this note is to explain the origin of this in terms of basic tokamak equilibrium concepts, namely the effects on the familiar $'s-\alpha'$ stability diagram of the Shafranov shift, $\Delta'(r)$, plasma shaping and 'favourable average curvature'.

The stability against high-*n* ballooning modes is usually investigated in full toroidal geometry, but with the stability boundaries being described in terms of a normalised pressure gradient parameter, $'\alpha'$, and plasma current density, $j$, or equivalently the magnetic shear, $s$, local to the magnetic surface being analysed. However this description hides other potential equilibrium dependencies such as on the Shafranov shift, $\Delta'(r)$, which is a consequence of the global pressure profile, i.e. the plasma beta, $\beta$, and the surface shape (e.g. ellipticity, $\kappa$). A role for the Shafranov shift has been proposed [11, 121], but it is not readily separable from other potential causes. In order to understand these aspects it is useful to develop a model ballooning equation that explicitly contains such parameters, so we need to consider an analytic, tokamak equilibrium. We take the large aspect ratio $\varepsilon = r/R \ll 1$, Shafranov tokamak equilibrium with $\beta \sim \varepsilon^2$ and approximately circular magnetic surfaces, $r = \text{constant}$, albeit displaced by the Shafranov shift, $\Delta'(r)$ with some weak shaping, in particular ellipticity, parameterised by a quantity $E(r)$, where the ellipticity $\kappa(r) = (E(r)+r)/(E(r)-r)$ [13]. The familiar $s-\alpha$ ballooning equation [14] corresponds to taking the limit $\varepsilon \to 0$, $\beta \to 0$, i.e. $\Delta'(r) \to 0$ of this equilibrium, while assuming a steep pressure gradient with $rd(\ell np/dr) \sim \varepsilon^{-1}$ exists in a narrow region, of width $\delta r$, in the vicinity of the surface under consideration, so that the parameter $\alpha = -2Rq^2(dp/dr)/B^2 \sim 0(1)$, where $q$ is the safety factor. Since the pressure gradient is only large in a narrow radial region, $\delta r \ll r$, one can consistently assume concentric circular magnetic surfaces [15].

In order to investigate the effect of $\beta$ on ballooning stability we consider a somewhat different modification of the Shafranov equilibrium in which the pressure gradient is enhanced globally, rather than locally. We also consider the limit of small magnetic shear, which simplifies the analysis by allowing a two-scale approach [16], but clearly shows the impact of the effects of finite $\beta$ on ballooning stability. An optimal ordering is adopted that allows competition between $\alpha$ and $\Delta'$ and also with the effects of mild plasma shaping and favourable average curvature due to retaining terms of $0(r/R_0)$:

$$\alpha \sim \Delta' \sim \lambda; \quad s \sim E \sim E' \sim \lambda^2; \quad r/R_0 \sim \lambda^3, \tag{1}$$

This allows the surfaces to remain circular in leading order, but to be self-consistent they must have some ellipticity at a level that is driven by $\beta$. We also allow for the possibility of an imposed ellipticity at the plasma boundary, parameterised by $E(a)$, at a comparable magnitude. Higher harmonic shaping such as triangularity is found not to contribute at low magnetic shear. The modified ballooning equation contains the parameters $\varepsilon = a/R_0$, $\Delta'$ and $E$, in addition to $s$ and $\alpha$, with $\Delta'$ and $E$ involving the effects of $\beta$ so that we can explicitly examine the effect of $\beta$ on ballooning stability.

## 2. MHD Ballooning Stability at Low Shear

The high-$n$, ideal MHD ballooning equation in general geometry is [17, 18]

$$\mathbf{B}.\nabla\left(\frac{|\nabla S|^2}{B^2}\mathbf{B}.\nabla\right)\varphi + 2\frac{\partial p}{\partial \psi}\frac{\mathbf{B}\times\nabla S.\mathbf{\kappa}}{B^2}\varphi = 0, \quad \text{where} \quad \mathbf{\kappa} = \frac{\mathbf{B}\times(\nabla(p+B^2/2)\times\mathbf{B})}{B^4} \tag{2}$$

is the curvature, $\psi$ the poloidal flux and $S$ is the eikonal, $S = \zeta - q\theta - \int kdr$, with $k$ a radial wave-number [19]. We use non-orthogonal straight field line co-ordinates $r$, $\theta$, $\zeta$ with Jacobian $J = rR^2/R_0$ [8] and express the equilibrium magnetic field as $\mathbf{B} = R_0 B_0 (g(r)\nabla\zeta + f(r)\nabla\zeta \times \nabla r)$, so that $q = rg/R_0 f$, with $r$ the magnetic surface label. In this co-ordinate system eqn. (2) becomes

$$\frac{1}{J}\frac{\partial}{\partial\theta}\left(\frac{|\nabla S|^2}{B^2}\frac{1}{J}\frac{\partial}{\partial\theta}\right)\varphi + 2\frac{\partial p}{\partial r}\left(\frac{q}{rB_0 g}\right)^3 \frac{\nabla(p+B^2/2)\times\mathbf{B}\cdot\nabla S}{B^4}\varphi = 0. \quad (3)$$

Here

$$|\nabla S|^2 = \frac{1}{R^2} + \frac{q^2}{r^2}\left(|r\nabla\theta|^2 + s^2(\theta-\theta_0)^2|\nabla r|^2 + 2s(\theta-\theta_0)r\nabla r\cdot\nabla\theta\right), \quad (4)$$

with $s = (r/q)dq/dr$ the magnetic shear and where we have introduced the ballooning angle, $k = -q'\theta_0$, in place of the radial wave-number $k$. We can express

$$\frac{\nabla(p+B^2/2)\times\mathbf{B}\cdot\nabla S}{B^4} = \frac{IR_0 q}{rR^2 B^2}(\kappa_r - s(\theta-\theta_0)\kappa_\theta), \quad (5)$$

where

$$\kappa_r = \frac{1}{B^2}\frac{\partial}{\partial r}(p + B^2/2), \quad \kappa_\theta = \frac{1}{B^2}\frac{\partial}{r\partial\theta}(p + B^2/2) \quad (6)$$

With the aid of eqns. (4)-(6), eqn. (3) can be written in the form

$$\frac{d}{d\theta}\left(F(\theta)\frac{d}{d\theta}\varphi\right) - \alpha G(\theta)\varphi = 0 \quad (7)$$

where

$$F(\theta) = \frac{r^2}{q^2}|\nabla S|^2; \quad G(\theta) = \left(\frac{R}{gR_0}\right)^2 (R_0\kappa_r - s(\theta-\theta_0)R_0\kappa_\theta) \quad (8)$$

At this point we introduce the ordering (1), the two scales, $\theta, u = s\theta$ [16], so that $\partial/\partial\theta \to (\partial/\partial\theta + \lambda^2 s\partial/\partial u)$ and expand

$$F = F_0 + \lambda F_1 + \lambda^2 F_2 + ....; \quad G = G_0 + \lambda G_1 + \lambda^2 G_2 + \lambda^3 G_3 + ..; \quad \varphi = \varphi_0 + \lambda\varphi_1 + \lambda^2\varphi_2 + \lambda^3\varphi_3 + \lambda^4\varphi_4 + .... \quad (9)$$

In $0(\lambda^0)$ we find

$$\frac{\partial}{\partial\theta}\left(F_0 \frac{\partial}{\partial\theta}\varphi_0\right) = 0 \quad (10)$$

so that $\varphi_0 = \varphi_0(u)$. In next order

$$\frac{\partial}{\partial \theta}\left(F_0 \frac{\partial}{\partial \theta}\varphi_1\right) = \alpha G_0 \varphi_0 \quad , \tag{11}$$

while in $0(\lambda^2)$

$$\frac{\partial}{\partial \theta}\left(F_0 \frac{\partial}{\partial \theta}\varphi_2\right) + \frac{\partial}{\partial \theta}\left(F_1 \frac{\partial}{\partial \theta}\varphi_1\right) + s\frac{\partial}{\partial \theta}\left(F_0 \frac{\partial}{\partial u}\varphi_0\right) = \alpha(G_0 \varphi_1 + G_1 \varphi_0). \tag{12}$$

In next order

$$\frac{\partial}{\partial \theta}\left(F_0 \frac{\partial}{\partial \theta}\varphi_3\right) + \frac{\partial}{\partial \theta}\left(F_1 \frac{\partial}{\partial \theta}\varphi_2\right) + \frac{\partial}{\partial \theta}\left(F_2 \frac{\partial}{\partial \theta}\varphi_1\right) + s\frac{\partial}{\partial \theta}\left(F_0 \frac{\partial}{\partial u}\varphi_1\right) + s\frac{\partial}{\partial u}\left(F_0 \frac{\partial}{\partial \theta}\varphi_1\right)$$

$$+ s\frac{\partial}{\partial \theta}\left(F_1 \frac{\partial}{\partial u}\varphi_0\right) = \alpha(G_0 \varphi_2 + G_1 \varphi_1 + G_2 \varphi_0). \tag{13}$$

Finally, in the $0(\lambda^4)$ equation, we annihilate the term in $\varphi_4$ by the operation $\langle ...\rangle = \oint d\theta(...)/2\pi$, to obtain

$$s\frac{\partial}{\partial u}\left\langle F_0 \frac{\partial}{\partial \theta}\varphi_2\right\rangle + s\frac{\partial}{\partial u}\left\langle F_1 \frac{\partial}{\partial \theta}\varphi_1\right\rangle + s^2 \frac{\partial}{\partial u}\left(F_0 \frac{\partial}{\partial u}\varphi_0\right)$$

$$= \alpha(\langle G_0 \varphi_3\rangle + \langle G_1 \varphi_2\rangle + \langle G_2 \varphi_1\rangle + \langle G_3\rangle \varphi_0) \tag{14}$$

Since we shall see that $F_0 = F_0(u)$, the first term on the left hand side vanishes and we only require $F(\theta)$ to order $\lambda^2$; furthermore, although we need to expand $G(\theta)$ to $0(\lambda^3)$, we only need retain that part of $G_3$ that is independent of periodic terms in $\theta$. Similarly, since we shall find that $\varphi_1$ contains only the $\cos\theta$ and $\sin\theta$ harmonics, we only need retain the same harmonics when calculating $G_2$.

### 3. Equilibrium Quantities for the Ballooning Equation

It remains to evaluate the geometrical quantities in eqn. (7). To do this we use the Shafranov equilibrium, expressed in co-ordinates $r'$, $\omega$, $\zeta$ through the representation [13]

$$R = R_0 - \lambda^3 r' \cos\omega - \lambda^4 \Delta(r') + \lambda^5 E(r')\cos\omega + \lambda^6 T(r')\cos 2\omega + \lambda^7 P(r')\cos\omega,$$
$$Z = \lambda^3 r' \sin\omega + \lambda^5 E(r')\sin\omega + \lambda^6 T(r')\sin 2\omega - \lambda^7 P(r')\sin\omega. \tag{15}$$

We note that we have measured the angle $\omega$ from the inboard side of the tokamak. Although the surfaces are taken to be circular one finds that the equilibrium pressure forces a small amount of ellipticity, parameterised by $E(r')$, and triangularity, $T(r')$, which can be removed at a given surface by applying external shaping but necessitates a local gradient and curvature of $E(r')$ and $T(r')$. The function $P(r')$ merely allows one to re-label surfaces; for the moment we have differentiated between the radial co-ordinates $r$ and $r'$ but we shall choose the function $P(r')$ later such that they can be identified and will from now on ignore the distinction.

We write $p(r) = \lambda^4 p_4(r) + ....; \quad g(r) = \left(1 + \lambda^4 g_4(r) + ....\right)$, since $p'_4 \sim \alpha(r/R_0) \sim \lambda^4$. Substitution of expansion (15) into the Grad-Shafranov equation yields equations for $p_4(r)$, $\Delta'(r)$ and $E(r)$ [13]:

$$\frac{p'_4}{B_0^2} + g'_4 + \frac{1}{R_0^2 q}\left(\frac{r^2}{q}\right)' = 0, \tag{16}$$

$$r\Delta'' + \left(2\left(\frac{r^2}{q}\right)'\frac{q}{r} - 1\right)\Delta' - \frac{2q}{R_0}\left(\frac{r^2}{q}\right)' - \frac{r}{R_0} - 2R_0 q^2 g'_4 = 0 \tag{17}$$

$$rE'' + \left(2\left(\frac{r^2}{q}\right)'\frac{q}{r} - 1\right)E' - \frac{3}{r}E + 3\left(\frac{r^2}{q}\right)'\frac{q}{2r}\Delta'^2 - \frac{3}{2}\Delta'^2 + \frac{3}{2}r\Delta'\Delta'' = 0 \tag{18}$$

where we ignore terms small in $\lambda$. We can integrate eqn. (17) to obtain $\Delta'(r)$. Since our ballooning analysis is applied near the plasma edge we effectively require $\Delta'(a)$ which is given by

$$\Delta'(a) = \frac{a}{R_0}\left(\beta_{pol} + \frac{l_i(a)}{2}\right) \tag{19}$$

where $\beta_{pol}$ is the poloidal $\beta$ of the plasma and $l_i$ is the internal inductance per unit length of the plasma column. Thus we see that $\Delta'$ is a parameter representing the *global* plasma $\beta$ and is distinct from the parameter $\alpha$, which only represents the *local* pressure gradient. For more precise calculations we should use

$$\Delta'(r) = \frac{r}{R_0}\left(\beta_{pol} + \frac{l_i(r)}{2}\right); \quad \beta_{pol}(r) = \frac{2}{B_\theta^2}\left(\frac{2}{r^2}\int_0^r prdr - p(r)\right), \quad l_i(r) = \frac{2}{r^2 B_\theta^2}\int_0^r B_\theta^2 rdr \quad, \tag{20}$$

but $l_i$ is smaller than $\beta_{pol}$ with our ordering (1). From eqn. (15) we can compute the Jacobian for this co-ordinate system:

$$\hat{j} = \frac{\partial R}{\partial \omega}\frac{\partial Z}{\partial r} - \frac{\partial R}{\partial r}\frac{\partial Z}{\partial \omega} \tag{21}$$

The straight field line angle $\theta$ is then obtained from [19]

$$\theta = 2\pi \int_0^\omega (R_0/R)\hat{J}d\omega / \oint (R_0/R)\hat{J}d\omega. \qquad (22)$$

The radial co-ordinate $r$ is defined as [19]:

$$r^2 = 2\int_0^r dr' \oint d\omega \hat{J} R_0/R \qquad (23)$$

Inserting the expansion into $(R_0/R)\hat{J}$ and expanding in $\varepsilon$ this serves to determine $P(r)$ as:

$$P(r) = \frac{1}{8}\frac{r^3}{R_0^2} + \frac{1}{2}\frac{r\Delta}{R_0} - \frac{1}{2}\frac{E^2}{r} \qquad (24)$$

Similarly, with the aid of eqn. (22), we can calculate $\theta(\omega)$ from eqns. (15), (21) and (22) and, on inverting the expression, we obtain

$$\omega = \theta - \Delta'\sin\theta + \frac{1}{2}\left(\Delta'^2 - \frac{E}{r} + E'\right)\sin 2\theta - \frac{r}{R_0}\sin\theta + \frac{\Delta'}{4}\left(\frac{\Delta'^2}{2} - 5\frac{E}{r} + E'\right)\sin\theta$$
$$- \left[\frac{3\Delta'}{4}\left(\frac{\Delta'^2}{2} - \frac{E}{r} + E'\right) - \frac{T'}{3}\right]\sin 3\theta \qquad (25)$$

To compute $|\nabla r|^2$, $r\nabla r.\nabla\theta$ and $r^2|\nabla\theta^2|$ we substitute the expansion (25) into eqn. (15) and form $\nabla R$ and $\nabla Z$ in terms of $\nabla r$ and $r\nabla\theta$. Inverting these expressions one can readily form $|\nabla r|^2$, $r\nabla r.\nabla\theta$ and $r^2|\nabla\theta^2|$ since $\nabla R$ and $\nabla Z$ are orthogonal unit vectors. After some lengthy algebra one obtains, to $0(\lambda^2)$,

$$|\nabla r|^2 = 1 - 2\Delta'\cos\theta + \frac{\Delta'^2}{2} + \left(\frac{5}{2}\Delta'^2 + 2E'\right)\cos 2\theta \qquad (26)$$

$$r^2|\nabla\theta|^2 = 1 + 2\Delta'\cos\theta + \frac{(r\Delta'')^2}{2} + r\Delta'\Delta'' + 2\Delta'^2 - \left(\Delta'^2 + r\Delta'\Delta'' + \frac{(r\Delta'')^2}{2} + 2E'\right)\cos 2\theta \qquad (27)$$

$$r\nabla r.\nabla\theta = (\Delta' + r\Delta'')\sin\theta - \frac{1}{2}\left(\Delta'^2 + r\Delta'\Delta'' + 3\frac{E}{r} + E' + rE''\right)\sin 2\theta. \qquad (28)$$

It remains to replace $r\Delta''$ and $rE''$ from eqns. (16) - (18). These are given by:

$$r\Delta'' = \alpha + (2s-3)\Delta' + \frac{r}{R_0}; \quad rE'' = -3\left(\frac{\Delta'^2}{2} + \frac{\alpha\Delta'}{2} + E' - \frac{E}{r}\right). \qquad (29)$$

to $O(\lambda^3)$, where we have substituted for $g'_4$ in favour of $p'_4$ from eqn. (16) and expressed $p'_4$ in terms of $\alpha$. Thus $r^2|\nabla\theta^2|$ and $r\nabla r.\nabla\theta$ are modified and when substituted in eqn. (4) we obtain

$$F(\theta) = \frac{r^2}{q^2}|\nabla S|^2 = 1 + 2\Delta'\cos\theta + \frac{7}{2}\Delta'^2 - 2\alpha\Delta' + \frac{\alpha^2}{2} - \left(\frac{5}{2}\Delta'^2 - 2\alpha\Delta' + \frac{\alpha^2}{2} + 2E'\right)\cos 2\theta$$
$$+ 2s(\theta - \theta_0)\theta\left[(\alpha - 2\Delta')\sin\theta + \left(2\Delta'^2 - \frac{3}{4}\alpha\Delta' - 3\frac{E}{r} + E'\right)\sin 2\theta\right]$$
$$+ s^2(\theta - \theta_0)^2\left[1 - 2\Delta'\cos\theta + \frac{\Delta'^2}{2} + \left(\frac{5}{2}\Delta'^2 + 2E'\right)\cos 2\theta\right]$$

(30)

We now evaluate $\kappa_r$ and $\kappa_\theta$ as defined in eqn. (6). Thus

$$R_0\kappa_r = -\frac{R_0}{2R^2}\frac{\partial}{\partial r}(R^2) + \frac{p'}{B_0^2}\left(\frac{R^2}{R_0^2} - 1\right) + R_0\left[\frac{p'_4}{B_0^2} + g'_4 + \frac{(f^2)'}{2}|\nabla r|^2\right],$$

(31)

on using eqns. (16) and (26), so that

$$\left(\frac{R^2}{R_0^2}\right)R_0\kappa_r \cong -\frac{R_0}{2}\frac{\partial}{\partial r}\left(\frac{R^2}{R_0^2}\right) - \frac{r}{R_0 q^2}$$

(32)

as required for forming $G(\theta)$. Likewise

$$\left(\frac{R^2}{R_0^2}\right)R_0\kappa_\theta = -\frac{R_0}{2r}\frac{\partial}{\partial\theta}\left(\frac{R^2}{R_0^2}\right)$$

(33)

Expressing $R^2$ as a function of $r$ and $\theta$ using eqns. (15) and (25), we obtain

$$\left(\frac{R}{R_0}\right)^2 = 1 - \frac{2r}{R_0}\cos\theta - \left(\frac{2\Delta}{R_0} + \frac{r\Delta'}{R_0}\right) + \frac{r\Delta'}{R_0}\cos 2\theta + \frac{r}{2R_0}\left(\frac{3}{2}\Delta'^2 + \frac{3E}{r} + E'\right)\cos\theta - \frac{r^2}{2R_0^2} + \ldots\ldots \quad (34)$$

where, as mentioned earlier, we only retain $\cos\theta$ and $\sin\theta$ harmonics in $O(\lambda^5)$ and constant terms in $O(\lambda^6)$ as these produce the required contributions to $G_2$ and $G_3$. It is interesting to note that neither $P$ nor $T$ (i.e. triangularity) contribute to $G_3$. Calculating $G(\theta)$ using eqns. (32), (33) and (34) and recalling eqn. (29) for $\Delta''$ and $E''$, we obtain

$$G(\theta) = \cos\theta + s(\theta - \theta_0)\sin\theta + \frac{\alpha}{2} - \frac{1}{2}(\alpha - 2\Delta' + 2s\Delta')\cos 2\theta - s(\theta - \theta_0)\Delta'\sin 2\theta$$
$$+ \frac{1}{4}\left(\frac{9}{2}\Delta'^2 - \frac{3\alpha\Delta'}{2} - E' - \frac{3E}{r}\right)\cos\theta - \frac{s(\theta - \theta_0)}{4}\sin\theta\left(\frac{3}{2}\Delta'^2 + \frac{3E}{r} + E'\right) + \frac{r}{R_0}\left(1 - \frac{1}{q^2}\right) + s\Delta' \tag{35}$$

### 4. The 'Averaged' Ballooning Equation

We are now in a position to develop eqn. (14) for $\varphi_0$ which determines ideal MHD ballooning stability at low magnetic shear. With the substitution $s(\theta - \theta_0) \to u$ in eqns. (30) and (35) we can identify the quantities $F_0, F_1, F_2, G_1, G_2$ and $G_3$. In particular, $F_0 = (1 + u^2)$, which is indeed independent of $\theta$. It is also convenient to introduce the notation

$$F(\theta) = F_0 + \lambda\{f_1^c \cos\theta + uf_1^s \sin\theta\} + \lambda^2\{\bar{f}_2 + f_2^c \cos 2\theta + uf_2^s \sin 2\theta\} + \ldots$$
$$G(\theta) = \cos\theta + u\sin\theta + \lambda\left\{\frac{\alpha}{2} + g_1^c \cos 2\theta + ug_1^s \sin 2\theta\right\} + \lambda^2\{g_1^c \cos\theta + ug_1^s \sin\theta + \ldots\} + \lambda^3\{\bar{g}_3 + \ldots\} + \ldots \tag{36}$$

We can readily evaluate $\langle G_3\rangle \varphi_0(u)$:

$$\langle G_3\rangle \varphi_0(u) = \bar{g}_3 \tag{37}$$

Equation (11) then yields

$$\varphi_1 = -\alpha \frac{\varphi_0(u)}{F_0}(\cos\theta + u\sin\theta) \quad , \tag{38}$$

from which we can evaluate $s\langle F_1 \partial\varphi_1/\partial\theta\rangle$ and $\langle G_2\varphi_1\rangle$

$$s\left\langle F_1\frac{\partial\varphi_1}{\partial\theta}\right\rangle = \frac{\alpha s}{2F_0}u(f_1^s - f_1^c)\varphi_0(u), \qquad \langle G_2\varphi_1\rangle = -\frac{\alpha}{2F_0}(g_2^c + u^2 g_2^s)\varphi_0(u) \tag{39}$$

From eqn. (12) we learn

$$F_0\frac{\partial\varphi_2}{\partial\theta} + F_1\frac{\partial\varphi_1}{\partial\theta} + sF_0\frac{\partial\varphi_0}{\partial u} = \frac{\alpha\varphi_0(u)}{2}\left\{-\frac{\alpha}{2F_0}\left((1-u^2)\sin 2\theta - 2u\cos 2\theta\right) + g_1^c \sin 2\theta - ug_1^s \cos 2\theta\right\} + C(u)$$

$$\tag{40}$$

Periodicity of $\varphi_2 \Rightarrow$

$$C(u) = \left\langle F_1 \frac{\partial \varphi_1}{\partial \theta} \right\rangle + sF_0 \frac{\partial \varphi_0(u)}{\partial u}. \tag{41}$$

$$F_0 \frac{\partial \varphi_2}{\partial \theta} = -\frac{\alpha \varphi_0(u)}{2F_0}\left\{(f_1^c - u^2 f_1^s)\sin 2\theta - u(f_1^c + f_1^s)\cos 2\theta\right\}$$
$$+ \frac{\alpha \varphi_0(u)}{2}\left\{-\frac{\alpha}{2F_0}\left((1-u^2)\sin 2\theta - 2u\cos 2\theta\right) + g_1^c \sin 2\theta - ug_1^s \cos 2\theta\right\} \tag{42}$$

By integrating by parts we can then evaluate $\langle G_1 \varphi_2 \rangle$:

$$\langle G_1 \varphi_2 \rangle = \frac{\alpha^2}{8F_0^2}\left\{g_1^c(f_1^c - u^2 f_1^s) + u^2 g_1^s(f_1^s + f_1^c) + \frac{\alpha}{2}\left[(1-u^2)g_1^c + 2u^2 g_1^s\right] - F_0\left[(g_1^c)^2 + (ug_1^s)^2\right]\right\}\varphi_0(u) \tag{43}$$

Finally, we can evaluate $\langle G_0 \varphi_3 \rangle$ by integrating twice by parts and using eqn. (13): this generates a plethora of terms:

$$\langle G_0 \varphi_3 \rangle = \left\langle \frac{G_0}{F_0}\left[\frac{\partial}{\partial \theta}\left(F_1 \frac{\partial}{\partial \theta}\varphi_2\right) + \frac{\partial}{\partial \theta}\left(F_2 \frac{\partial}{\partial \theta}\varphi_1\right) + s\frac{\partial}{\partial \theta}\left(F_0 \frac{\partial}{\partial u}\varphi_1\right) + s\frac{\partial}{\partial u}\left(F_0 \frac{\partial}{\partial \theta}\varphi_1\right) + s\frac{\partial}{\partial \theta}\left(F_1 \frac{\partial}{\partial u}\varphi_0\right)\right]\right\rangle$$
$$- \alpha \left\langle \frac{G_0}{F_0}(G_0 \varphi_2 + G_1 \varphi_1 + G_2 \varphi_0)\right\rangle \tag{44}$$

These are evaluated in the appendix and given in eqn. (A.9).

The results (37), (39), (43) and (A.9) provide all the information needed to complete eqn. (14) for $\varphi_0(u)$. The development of this equation is also described in more detail in the appendix. The result is

$$s^2 \frac{d}{du}\left(F_0 \frac{d}{du}\varphi_0\right) = \left(A_0 + \frac{A_1}{F_0} + \frac{A_2}{F_0^2}\right)\varphi_0 \tag{45}$$

where

$$A_0 = \alpha\left[\frac{r}{R_0}\left(1 - \frac{1}{q^2}\right) + \frac{9}{8}\alpha\Delta'^2 + \frac{3}{4}\alpha\left(\frac{E}{r} + E'\right)\right] \tag{46}$$

$$A_1 = \alpha\left[\frac{9}{8}\alpha^2\Delta' - \frac{9}{2}\alpha\Delta'^2 + 3\alpha\left(\frac{E}{r} - E'\right) - 4s\Delta'\right] \tag{47}$$

$$A_2 = \alpha\left[\frac{3}{8}\alpha^3 - \frac{9}{4}\alpha^2\Delta' + \frac{9}{2}\alpha\Delta'^2 - 3\alpha\left(\frac{E}{r} - E'\right) + 2s(4\Delta' - \alpha)\right] \tag{48}$$

The form of this equation has been confirmed by using computer algebra [20].

It is interesting to consider the large $u$ limit:

$$s^2 \frac{d}{du}\left(F_0 \frac{d}{du}\varphi_0\right) = \alpha\left[\frac{r}{R_0}\left(1-\frac{1}{q^2}\right) + \frac{9}{8}\alpha\Delta'^2 + \frac{3}{4}\alpha\left(\frac{E}{r}+E'\right)\right]\varphi_0, \tag{49}$$

yielding the Mercier criterion [21]:

$$\frac{s^2}{4} + \alpha\left[\frac{r}{R_0}\left(1-\frac{1}{q^2}\right) + \frac{9}{8}\alpha\Delta'^2 + \frac{3}{4}\alpha\left(\frac{E}{r}+E'\right)\right] > 0, \tag{50}$$

which we can write in the standard form:

$$\frac{1}{4} + D_M > 0; \quad D_M = \frac{\alpha}{s^2}\left[\frac{r}{R_0}\left(1-\frac{1}{q^2}\right) + \frac{9}{8}\alpha\Delta'^2 + \frac{3}{4}\alpha\left(\frac{E}{r}+E'\right)\right]. \tag{51}$$

## 5. Ideal MHD Ballooning Stability

We now investigate the marginal stability curves corresponding to solutions of eqn. (45) that vanish as $|u| \to \infty$, determining the impact of $\beta$ (through its impact on $\Delta'$ and $E$) and $a/R_0$ on the $s-\alpha$ diagram at low magnetic shear. Several influences can affect the stability with respect to the $s-\alpha$ diagram: the role of finite aspect ratio (the first term in $A_0$, $d = (r/R_0)(1-q^{-2})$); the role of the Shafranov shift $\Delta'$ as $\beta$ increases; and finally the effect of ellipticity through $E$, which has a direct effect through the shaping but also through $E'$ which itself responds to $\beta$. A fully self-consistent treatment requires all these effects to be included, but it is instructive to consider their effects sequentially.

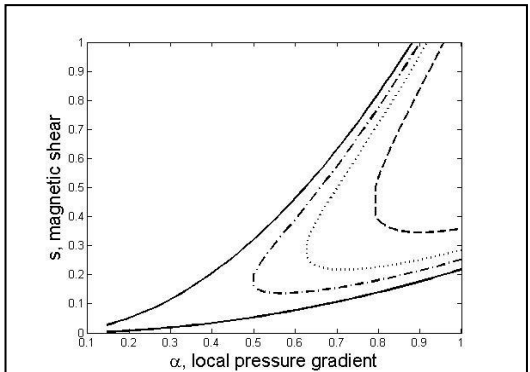

Fig. 1: The effect of finite aspect ratio on the $s-\alpha$ stability diagram. The solid line shows $a/R_0 = 0$, the dash-dot line $a/R_0 = 0.025$, the dotted line $a/R_0 = 0.05$, and the dashed line $a/R_0 = 0.1$ when $q = 3$.

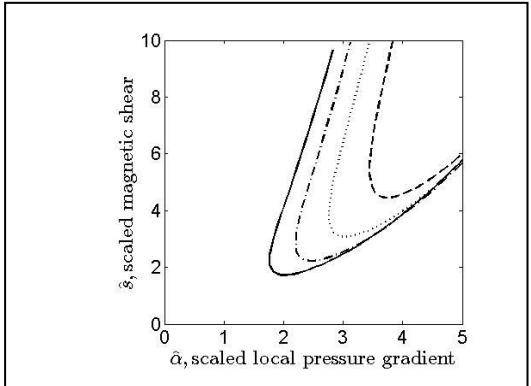

Fig. 2: The effect of increasing $\Delta'$ on the scaled $s-\alpha$ stability diagram. The solid line shows $\hat{\Delta}'=0$, the dash-dot line shows $\hat{\Delta}'=0.3$, the dotted line shows $\hat{\Delta}'=0.6$, and the dashed line shows $\hat{\Delta}'=0.9$.

In Fig. 1 we show the effect of $a/R_0$ through $d$ in isolation, for several values of $a/R_0$ with a typical value for $q$ (i.e. $q=3$); thus stability is seen to increase rapidly with increasing $a/R_0$, relative to the standard $s-\alpha$ diagram. Note that, strictly, the analysis is only valid for the region $s \ll 1$. Having established the effect of $d$, we can effectively remove it as a parameter by scaling eqn. (45) so that it involves just four independent parameters, namely: $\alpha/d^{1/3}, \Delta'/d^{1/3}, E/d^{2/3}$ and $s/d^{2/3}$. The corresponding scaled quantity for $\beta$ is $\hat{\beta}=\beta_0/d^{4/3}$, where $\beta_0 = 2p(0)/B_0^2$, which we need when we calculate $\Delta'$ and $E$. We can thus plot stability curves parameterised by $\hat{\beta}$ through its impact on $\Delta'$ and $E'$ (for given values of $E(a)$) in a 'normalised $s-\alpha$ diagram' labelled by axes $\hat{\alpha}=\alpha/d^{1/3}$, and $\hat{s}=s/d^{2/3}$. However, before considering the complete problem we examine the effect of increasing the parameter $\hat{\Delta}'=\Delta'/d^{1/3}$ alone (one of the two outcomes of increasing $\hat{\beta}$), the results being shown in Fig. 2: again we observe a stabilising effect.

To address the complete problem we must first determine $E$ and $E'$, which involves a global solution for these quantities. In this ordering these are obtained from solutions of the equation

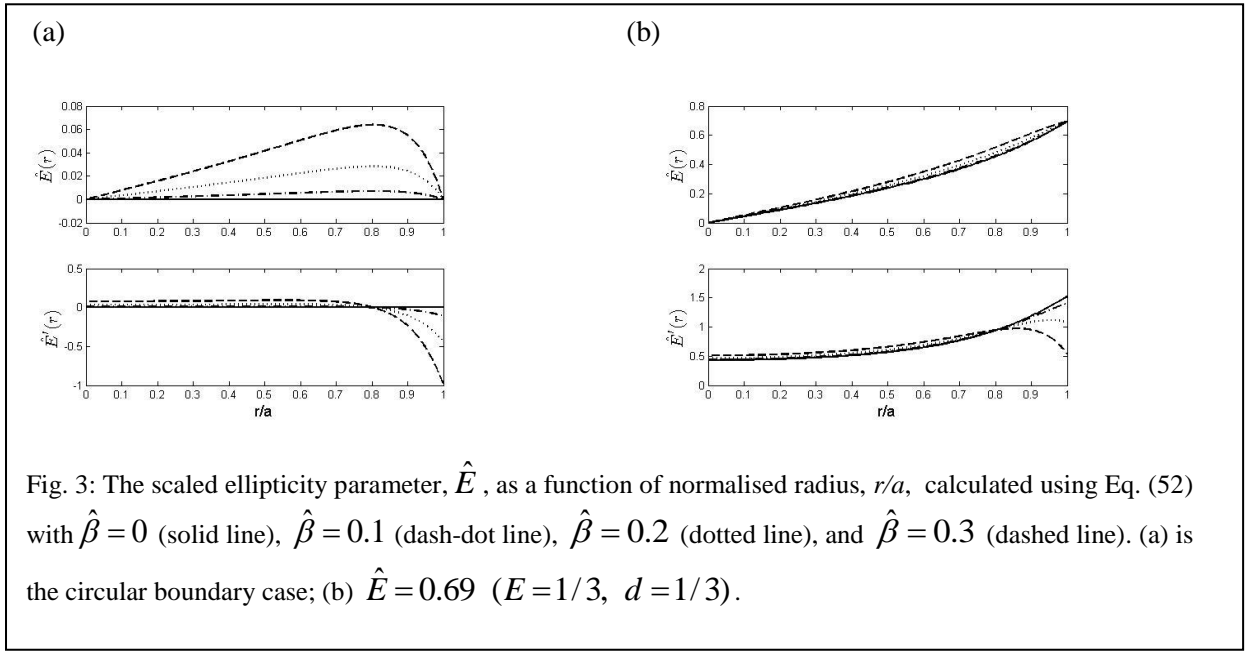

Fig. 3: The scaled ellipticity parameter, $\hat{E}$, as a function of normalised radius, $r/a$, calculated using Eq. (52) with $\hat{\beta}=0$ (solid line), $\hat{\beta}=0.1$ (dash-dot line), $\hat{\beta}=0.2$ (dotted line), and $\hat{\beta}=0.3$ (dashed line). (a) is the circular boundary case; (b) $\hat{E}=0.69$ ($E=1/3$, $d=1/3$).

$$E'' + \left(\frac{2q}{r^2}\left(\frac{r^2}{q}\right)' - \frac{1}{r}\right)E' - \frac{3}{r^2}E = -\frac{3q}{2r^2}\left(\frac{r^2}{q}\right)'\Delta'^2 + \frac{3}{2r}\Delta'^2 - \frac{3}{4}\frac{d\Delta'^2}{dr} = H(r) \qquad (52)$$

satisfying a boundary condition on $E(a)$, where

$$\Delta'(r) = -\frac{2R_0 q^2}{r^3 B_0^2}\int_0^r dr\, r^2 \frac{dp}{dr}. \qquad (53)$$

In our illustrative calculations we take simple global pressure and safety factor profiles of the form

$$p(r) = p(0)\left(1 - \left(\frac{r}{a}\right)^2\right); \qquad q(r) = \frac{q(0)}{1 - \left(\frac{r}{a}\right)^2 + \frac{1}{3}\left(\frac{r}{a}\right)^4} \qquad (54)$$

with $q(0) = 1$, so that $q(a) = 3$. In Fig.3 we show radial profiles of $\hat{E} = E/d^{2/3}$ for several values of $\hat{\beta}$: in Fig. 3(a) we choose $\hat{E}(a) = 0$, the circular boundary case as a reference, while in Fig. 3(b) we set $\hat{E}(a) = 0.69$, corresponding to a JET-like situation with $E(a) = 1/3$ and $d = 1/3$. The magnitude, and even the sign, of $\hat{E}'(a)$ is seen to vary with $\hat{\beta}$. While we observe that $\hat{E}'$ is negative near the edge, we note that the ellipticity, $\kappa$, is nevertheless an increasing function.

Using this information on $\hat{E}(a)$ and $\hat{E}'(a)$ and calculating $\hat{\Delta}'(a)$ from eqn. (53) as input, we solve the averaged ballooning equation (45). We must emphasize that we can treat $\hat{s}$ and $\hat{\alpha}$ as free,

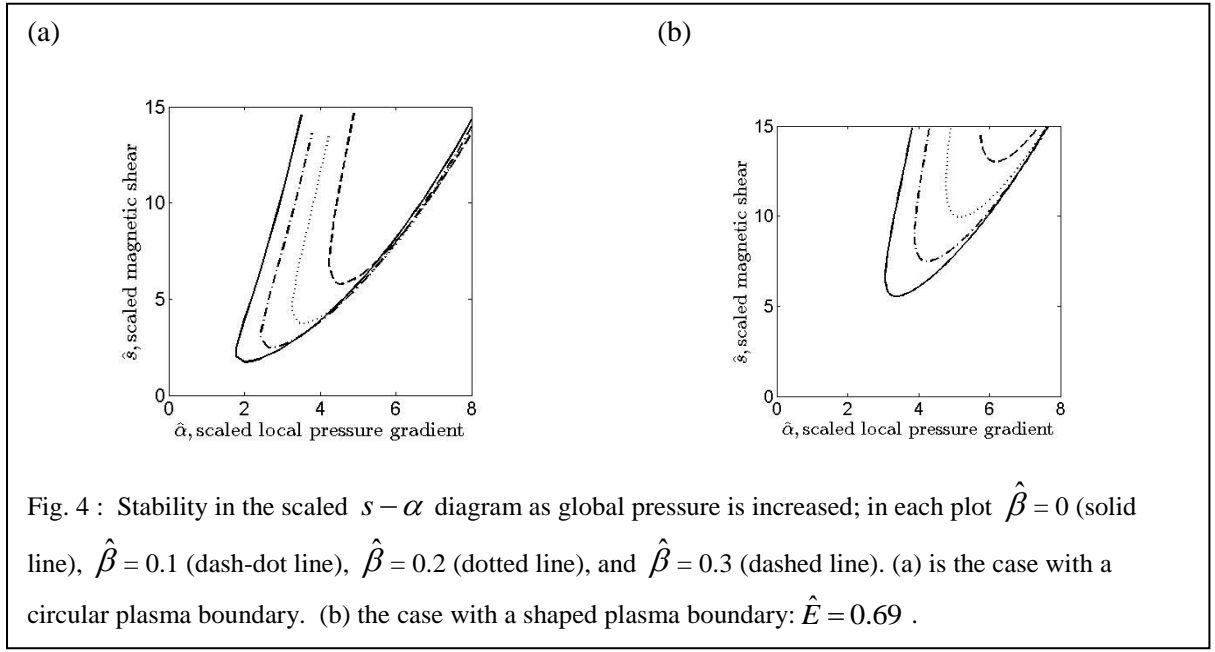

Fig. 4 : Stability in the scaled $s - \alpha$ diagram as global pressure is increased; in each plot $\hat{\beta} = 0$ (solid line), $\hat{\beta} = 0.1$ (dash-dot line), $\hat{\beta} = 0.2$ (dotted line), and $\hat{\beta} = 0.3$ (dashed line). (a) is the case with a circular plasma boundary. (b) the case with a shaped plasma boundary: $\hat{E} = 0.69$.

independent parameters on a given flux surface as the global equilibrium changes with increasing $\hat{\beta}$,

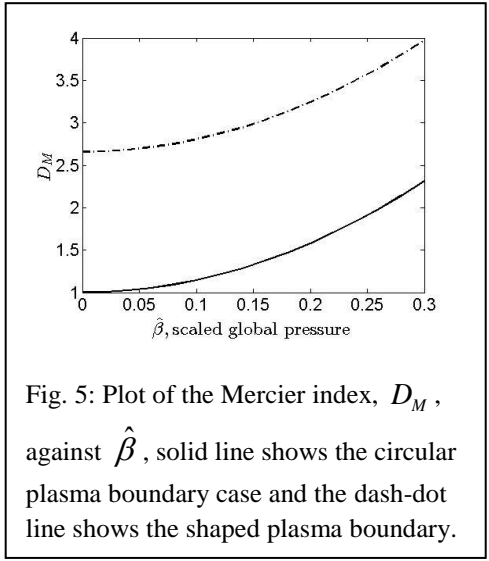

Fig. 5: Plot of the Mercier index, $D_M$, against $\hat{\beta}$, solid line shows the circular plasma boundary case and the dash-dot line shows the shaped plasma boundary.

since the necessary sharp gradients in these quantities can be considered to exist only over a localised region, without affecting the magnetic geometry of the underlying equilibrium [14, 15]. The curves of marginal stability in the $\hat{s} - \hat{\alpha}$ diagram corresponding to $r = a$ are shown in Fig. 4: Fig. 4(a) shows the impact of increasing $\hat{\beta}$ for the 'cylindrical' case $\hat{E}(a) = 0$, while Fig. 4(b) repeats this for the finite ellipticity case $\hat{E}(a) = 0.69$. We see that the stable regions increase as $\hat{\beta}$ increases for both cases and that a finite value of $\hat{E}(a)$ provides further stabilisation.

Finally, in Fig. 5 we show the dependence of $D_M$, related to Mercier stability, as a function of $\hat{\beta}$. Clearly it becomes yet more positive, precluding any question of Mercier instability.

## 6. Conclusions

The confinement properties of tokamaks depend on the H-mode pedestal characteristics and there is experimental evidence that these improve with plasma pressure. High-n ideal MHD ballooning modes, which may serve as an indicator for the effect of the more relevant kinetic ballooning modes, are believed to play a role in defining the pedestal properties. However, simple analysis based on the $s - \alpha$ stability diagram assumes that only the local pressure gradient, not the global pressure, or $\beta$, matters.

In this paper we have explored the impact of plasma $\beta$, as mediated through the Shafranov shift, $\Delta'$, for example, on stability. We employ a Shafranov equilibrium, but one in which the pressure is enhanced globally to allow $\beta$ to compete with $\alpha$, whereas in the $s - \alpha$ equilibrium calculation one only enhances $\alpha$ locally. This means that we also have to consider the effect of plasma ellipticity, parameterised by $E$, since it too responds to $\beta$. We also include the stabilising effects of favourable average curvature, proportional to $d = (a/R_0)(1 - q^{-2})$. In order to extract the essence of such effects we have considered the limit of low magnetic shear, $s$, (the region of the $s - \alpha$ diagram where the relative impact of $\Delta'$ and $d$ is greatest), allowing access to the second stability region for example. Furthermore the presence of the bootstrap current in the pedestal region tends to produce low $s$, making the calculation even more relevant. An added advantage of this region of parameter space is that a two-scale averaging process can be invoked to reduce the ideal MHD ballooning equation to a simpler form devoid of the usual poloidally periodic terms. An optimal ordering scheme ($\alpha \sim \Delta' \sim \lambda;\ s \sim E \sim \lambda^2;\ r/R_0 \sim \lambda^3$) for the quantities $\alpha, s, \Delta', a/R_0$ and $E$ has been introduced which ensures that all these effects on stability compete equally. This equilibrium information has been fed into the general high-$n$ MHD ballooning equation, which has then been processed order by order to generate the averaged equation (eqn.(45)), which appears in $O(\lambda^4)$. The equilibrium information required is reduced as a result of the averaging process: had we attempted to calculate the complete modified $s-\alpha$ diagram we would have needed to account for more poloidal harmonic structure in the higher orders of $\lambda$, including the presence of triangularity driven by $\beta$.

The averaged marginal MHD ballooning equation has been solved and the effects of $\Delta'$, $E$ and $d$ on the $s-\alpha$ diagram explored. There are thus six parameters to set: $s, \alpha, d, \Delta', E$ and $E'$. Since $E'$ plays a role we have to solve the global equilibrium equation for $E(r)$ to calculate this quantity. Examples of these solutions for given values of the imposed edge ellipticity parameter, $E(a)=0$ (the circular boundary case) and $E(a)=1/3$ (a JET-like case) for several values of $\beta$ are shown in Fig. 4.

Figure 1 demonstrates the effect of just including the effect of $d$, setting $\Delta'=E=E'=0$; clearly second stability access is rapidly opened up with increasing $a/R_0$. In order to explore the effects of $\beta$ through its impact on $\Delta'$ and $E$ it is convenient to reduce the number of independent parameters by a suitable scaling, introducing $\hat{\alpha}=\alpha/d^{1/3}$, $\hat{\Delta}'=\Delta'/d^{1/3}$, $\hat{E}=E/d^{2/3}$ and $\hat{s}=s/d^{2/3}$ (we also define $\hat{\beta}=\beta/d^{4/3}$, needed when calculating the radial profiles of $\hat{\Delta}'(r)$ and $E(r)$). In Fig.2 we examine the effect of just including the normalized Shafranov shift parameter, $\hat{\Delta}'$, alone, although this is not a consistent procedure; increasing $\hat{\Delta}'$ is seen to have a stabilizing effect. Finally we examine the full effect of $\beta$ through its self-consistent impact on $\hat{\Delta}'$, $E$ and $E'$. Figure 4 shows this to be stabilizing for the two cases considered: $E(a)=0, 1/3$. Finally, we note that increasing $\beta$ leads to increasingly Mercier stability, as shown in Fig. 5.

In summary we find that increasing $\beta$ has a stabilising effect on the $s-\alpha$ diagram through its impact on the Shafranov shift, $\Delta'$, and ellipticity parameter, $E$, providing a potential explanation for the experimental observations that the pedestal characteristics pertaining to tokamak confinement improve with increasing plasma pressure, even if the local pressure gradient is unaffected, and complementing studies with full MHD stability codes.

**Acknowledgments**

This work has received funding from the RCUK Energy Programme [grant EP/I501045]. For further information on this paper contact PublicationsManager@ccfe.ac.uk.

**Appendix: Details of the Derivation of the Averaged Ballooning Equation. (45).**

In this appendix we provide more detail on the derivation of eqn. (45). First we evaluate the quantity $\langle G_0 \varphi_3 \rangle$, given by eqn. (44):

$$\langle G_0 \varphi_3 \rangle = \left\langle \frac{G_0}{F_0} \left[ \frac{\partial}{\partial \theta}\left(F_1 \frac{\partial}{\partial \theta}\varphi_2\right) + \frac{\partial}{\partial \theta}\left(F_2 \frac{\partial}{\partial \theta}\varphi_1\right) + s\frac{\partial}{\partial \theta}\left(F_0 \frac{\partial}{\partial u}\varphi_1\right) + s\frac{\partial}{\partial u}\left(F_0 \frac{\partial}{\partial \theta}\varphi_1\right) + s\frac{\partial}{\partial \theta}\left(F_1 \frac{\partial}{\partial u}\varphi_0\right) \right] \right\rangle$$
$$- \alpha \left\langle \frac{G_0}{F_0}\left(G_0 \varphi_2 + G_1 \varphi_1 + G_2 \varphi_0\right) \right\rangle \quad \text{(A.1)}$$

term by term. For the first term on the right hand side, we obtain

$$\left\langle \frac{G_0}{F_0} \frac{\partial}{\partial \theta}\left( F_1 \frac{\partial}{\partial \theta} \varphi_2 \right)\right\rangle$$
$$= -\frac{\alpha}{8F_0^2}\left\{\left((f_1^c)^2 + (uf_1^s)^2\right) + \frac{\alpha}{2F_0}\left(f_1^c + u^2 f_1^s\right) - \left(f_1^c - u^2 f_1^s\right)g_1^c - u^2\left(f_1^s + f_1^c\right)g_1^s\right\}\varphi_0 \quad (A.2)$$

The second results in

$$\left\langle \frac{G_0}{F_0} \frac{\partial}{\partial \theta}\left( F_2 \frac{\partial}{\partial \theta} \varphi_1 \right)\right\rangle = \frac{\alpha}{4F_0^2}\left[2F_0 \bar{f}_2 - (1-u^2)f_2^c - 2u^2 f_2^s\right]\varphi_0 \quad (A.3)$$

The third and fourth are equal:

$$\left\langle \frac{G_0}{F_0} s \frac{\partial}{\partial \theta}\left( F_0 \frac{\partial}{\partial u} \varphi_1 \right)\right\rangle = \left\langle \frac{G_0}{F_0} s \frac{\partial}{\partial u}\left( F_0 \frac{\partial}{\partial \theta} \varphi_1 \right)\right\rangle = -\frac{s\alpha}{2F_0}\varphi_0 \quad (A.4)$$

while for the fifth,

$$\left\langle \frac{G_0}{F_0} s \frac{\partial}{\partial \theta}\left( F_1 \frac{\partial}{\partial u} \varphi_0 \right)\right\rangle = \frac{s}{2F_0} u\left(f_1^s - f_1^c\right)\frac{\partial}{\partial u}\varphi_0 \quad (A.5)$$

Turning to the terms involving $G$, we have

$$\alpha\left\langle \frac{G_0^2}{F_0} \varphi_2 \right\rangle = \frac{\alpha^2}{16F_0^2}\left[\left(f_1^c + u^2 f_1^s\right) - (1-u^2)g_1^c - 2u^2 g_1^s + \frac{\alpha F_0}{2}\right]\varphi_0 \, , \quad (A.6)$$

$$\alpha\left\langle \frac{G_0}{F_0} G_1 \varphi_1 \right\rangle = -\frac{\alpha^3}{4F_0}\varphi_0 - \frac{\alpha^2}{4F_0^2}\left[(1-u^2)g_1^c + 2u^2 g_1^s\right]\varphi_0 \quad (A.7)$$

and

$$\alpha\left\langle \frac{G_0}{F_0} G_2 \varphi_0 \right\rangle = \frac{\alpha}{2F_0}\left(g_2^c + u^2 g_2^s\right)\varphi_0 \quad . \quad (A.8)$$

Assembling these results, we finally have

$$\langle G_0\varphi_3\rangle = -\frac{\alpha}{8F_0^2}\left\{\left((f_1^c)^2+(uf_1^s)^2\right)+2\left[(1-u^2)f_2^c+2u^2f_2^s\right]+\alpha^2(f_1^c+u^2f_1^s)\right\}\varphi_0(u)$$
$$+\frac{\alpha}{8F_0^2}\left[(f_1^c-u^2f_1^s)g_1^c+(f_1^s+f_1^c)u^2g_1^s\right]\varphi_0(u)+\frac{5\alpha^2}{16F_0^2}\left[(1-u^2)g_1^c+2u^2g_1^s\right]\varphi_0(u)$$
$$-\frac{\alpha}{2F_0}(g_2^c+u^2g_2^s)\varphi_0(u)+\frac{7\alpha^2}{32F_0}\varphi_0(u)+\frac{\alpha}{2F_0}\bar{f}_2\varphi_0(u)-\frac{s\alpha}{F_0}\varphi_0(u)+\frac{su}{2F_0}(f_1^s-f_1^c)\frac{\partial}{\partial u}\varphi_0(u)$$
(A.9)

Now we turn to the derivation of eqn. (45). The results (37), (39), (43) and (A.9) provide all the information needed to complete eqn. (14) for $\varphi_0(u)$:

$$s^2\frac{d}{du}\left(F_0\frac{d}{du}\varphi_0\right) = \frac{\alpha}{F_0}\left[\frac{7}{32}\alpha^3+\frac{\alpha}{2}\bar{f}_2-s\alpha\right]\varphi_0 + \frac{\alpha^2}{4F_0^2}\left[g_1^c(f_1^c-u^2f_1^s)+ug_1^s(f_1^c+f_1^s)\right]\varphi_0$$
$$-\frac{\alpha^2}{8F_0^2}\left\{(uf_1^c)^2+(uf_1^s)^2+\alpha(f_1^c+u^2f_1^s)+2\left[(1-u^2)f_2^c+2u^2f_2^s\right]\right\}\varphi_0+\frac{s\alpha}{2}\frac{d}{du}\left(\frac{u}{F_0}(f_1^c-f_1^s)\right)\varphi_0$$
$$-\frac{\alpha^2}{8F_0}\left\{(g_1^c)^2+(ug_1^s)^2+4(g_2^c+u^2g_2^s)-\frac{3\alpha}{F_0}\left[(1-u^2)g_1^c+2u^2g_1^s\right]\right\}\varphi_0+\alpha\bar{g}_3\varphi_0.$$

(A.10)

It is convenient to isolate the u dependence of the coefficients above by substituting:

$$f_1^c=\bar{f}_1^c+u^2\hat{f}_1^c;\quad f_2^c=\bar{f}_2^c+u^2\hat{f}_2^c\quad\text{and}\quad \bar{f}_2=\bar{\bar{f}}_2+u^2\hat{\bar{f}}_2 \quad (A.11)$$

Then eqn. (A.10) can be rewritten as

$$s^2\frac{d}{du}\left(F_0\frac{d}{du}\varphi_0\right)=\left(A_0+\frac{A_1}{F_0}+\frac{A_2}{F_0^2}\right)\varphi_0 \quad (A.12)$$

Where

$$A_0=\alpha\bar{g}_3-\frac{\alpha^2}{8}(g_1^s)^2-\frac{\alpha^2}{2}g_2^s-\frac{s\alpha}{2}\hat{f}_1^c+\frac{\alpha^2}{4}\hat{f}_2^c-\frac{\alpha^2}{8}(\hat{f}_1^c)^2+\frac{\alpha^2}{4}g_1^s\hat{f}_1^c+\frac{\alpha^2}{2}\hat{\bar{f}}_2, \quad (A.13)$$

$$A_1 = \frac{3\alpha^3}{8}(2g_1^s - g_1^c) - \frac{\alpha^2}{8}\left((g_1^c)^2 - (g_1^s)^2\right) - \frac{\alpha^2}{2}(g_2^c - g_2^s) + \frac{\alpha^2}{4}\left[(g_1^c - g_1^s)(\hat{f}_1^c - f_1^s) + g_1^s(\hat{f}_1^c - \bar{f}_1^c)\right]$$

$$+ \alpha^2\left[\frac{7}{32}\alpha^2 + \frac{1}{2}(\bar{\bar{f}}_2 - \hat{\hat{f}}_2) - s\right] - \frac{s\alpha}{2}(f_1^s - \bar{f}_1^c + \hat{f}_1^c) - \frac{\alpha^2}{4}(2f_2^s - \bar{f}_2^c + 3\hat{f}_2^c)$$

$$- \frac{\alpha^3}{8}(\hat{f}_1^c + f_1^s) - \frac{\alpha^2}{8}\left[2\hat{f}_1^c(\bar{f}_1^c - \hat{f}_1^c) + (f_1^s)^2\right],$$

(A.14)

$$A_2 = \frac{3\alpha^3}{8}(2g_1^c - g_1^s) - s\alpha(\bar{f}_1^c - f_1^s - \hat{f}_1^c) - \frac{\alpha^2}{2}(\bar{f}_2^c - f_2^s - \hat{f}_2^c) - \frac{\alpha^3}{8}(\bar{f}_1^c - \hat{f}_1^c - f_1^s)$$

$$+ \frac{\alpha^2}{4}(g_1^c - g_1^s)(\bar{f}_1^c - \hat{f}_1^c + f_1^s) - \frac{\alpha^2}{8}\left[(\bar{f}_1^c - \hat{f}_1^c)^2 - (f_1^s)^2\right].$$

(A.15)

Recalling the definitions

$$f_1^c = \bar{f}_1^c + u^2\hat{f}_1^c = 2\Delta'(1 - u^2), \quad f_1^s = 2(\alpha - 2\Delta'), \quad \bar{f}_2 = \bar{\bar{f}}_2 + u^2\hat{\hat{f}}_2 = \frac{7}{2}\Delta'^2 - 2\alpha\Delta' + \frac{\alpha^2}{2} + \frac{\Delta'^2}{2}u^2,$$

$$f_2^c = \bar{f}_2^c + u^2\hat{f}_2^c = -\frac{5}{2}\Delta'^2 + 2\alpha\Delta' - \frac{\alpha^2}{2} - 2E' + \left(\frac{5}{2}\Delta'^2 + 2E'\right)u^2, \quad f_2^s = 4\Delta'^2 - \frac{3}{2}\alpha\Delta' - 6\frac{E}{r} + 2E';$$

$$g_1^c = -\frac{\alpha}{2} + \Delta', \quad g_1^s = -\Delta', \quad g_2^c = \frac{9}{8}\Delta'^2 - \frac{3}{8}\alpha\Delta' - \frac{3E}{4r} - \frac{E'}{4}, \quad g_2^s = -\frac{3}{8}\Delta'^2 - \frac{3E}{4r} - \frac{E'}{4}, \quad \bar{g}_3 = \frac{r}{R_0}\left(1 - \frac{1}{q^2}\right) + s\Delta'$$

(A.16)

we finally obtain

$$A_0 = \alpha\left[\frac{r}{R_0}\left(1 - \frac{1}{q^2}\right) + \frac{9}{8}\alpha\Delta'^2 + \frac{3}{4}\alpha\left(\frac{E}{r} + E'\right)\right]$$ (A.17)

$$A_1 = \alpha\left[\frac{9}{8}\alpha^2\Delta' - \frac{9}{2}\alpha\Delta'^2 + 3\alpha\left(\frac{E}{r} - E'\right) - 4s\Delta'\right]$$ (A.18)

$$A_2 = \alpha\left[\frac{3}{8}\alpha^3 - \frac{9}{4}\alpha^2\Delta' + \frac{9}{2}\alpha\Delta'^2 - 3\alpha\left(\frac{E}{r} - E'\right) + 2s(4\Delta' - \alpha)\right]$$ (A.19)

Equation (A.12), now with the coefficients (A.17) – (A.19), is the required equation for $\varphi_0(u)$, given as eqn. (45) - (49) in the main text.